# Designing and analyzing microwave photonic spectral domain filters based on transversal filtering with optical microcombs

**David J. Moss**[a, *]

[a]Optical Sciences Centre, Swinburne University of Technology, Hawthorn, VIC 3122, Australia

**Abstract**

Microwave transversal filters, which are implemented based on the transversal filter structure in digital signal processing, offer a high reconfigurability for achieving a variety of signal processing functions without changing hardware. When implemented using microwave photonic (MWP) technologies, also known as MWP transversal filters, they provide competitive advantages over their electrical counterparts, such as low loss, large operation bandwidth, and strong immunity to electromagnetic interference. Recent advances in high-performance optical microcombs provide compact and powerful multi-wavelength sources for MWP transversal filters that require a larger number of wavelength channels to achieve high performance, allowing for the demonstration of a diverse range of filter functions with improved performance and new features. Here, we present a comprehensive performance analysis for microcomb-based MWP spectral filters based on the transversal filter approach. First, we investigate the theoretical limitations in the filter spectral response induced by finite tap numbers. Next, we analyze the distortions in the filter spectral response resulting from experimental error sources. Finally, we assess the influence of input signal's bandwidth on the filtering errors. These results provide a valuable guide for the design and optimization of microcomb-based MWP transversal filters for a variety of applications.

## 1. Introduction

Microwave filters are widely used in telecommunication and radar systems to selectively pass or reject specific frequency components of microwave signals [1, 2]. Compared to conventional microwave filters that rely on electronic devices and hence suffer from limited operation bandwidths, microwave photonic (MWP) filters leverage optical filters with broad operation bandwidths to process high-bandwidth microwave signals that are modulated onto optical carriers. This can not only overcome the electrical bandwidth bottleneck, but also offer other attractive advantages including low loss, strong immunity to electromagnetic interference, fast tunability, and high flexibility in filter design [3, 4].

To implement MWP filters, there are primarily two approaches. The first approach involves directly mapping the response of optical filters onto the electrical domain. Examples of this approach include MWP filters based on passive optical filters (such as mirroring resonators [5-8]) and stimulated Brillouin scattering in waveguides [9-11]. The second approach involves synthesizing the desired filter response based on a digital filter structure known as the transversal filter structure [12-15], where weighted and progressively delayed replicas of the input microwave signal are generated at different wavelength channels and then combined upon photo-detection [16]. MWP filters implemented based on the transversal filter structure are commonly referred to as MWP transversal filters. These filters enable the realization of reconfigurable transfer functions that can be engineered to achieve a wide range of functions including signal generation [17-19], signal computing [20-25], and advanced adaptive and dynamic spectral filtering [14, 15, 26-29].

**Table I. Comparison of state-of-the-art MWP transversal filters. CW: continuous wave. AOTF: acousto-optic tunable filter. FBGs: fiber Bragg gratings. LPF: low-pass filter. BPF: band-pass filter. HPF: high-pass filter. BSF: band-stop filter. EO: electro-optic.**

| Source for generating multiple optical channels | Channel spacing (GHz)[a)] | Tap number | Filter shape | Year | Ref. |
|---|---|---|---|---|---|
| Discrete laser array | ~375 | 4 | BPF | 2003 | [30] |
| A broadband optical source + AOTF | ~725 – ~1500 | 2 | LPF | 2004 | [31] |
| A CW laser + FBGs array | ~375 | 2 | BPF | 1995 | [12] |
| A CW laser + FBGs array | N. A. [b)] | 16 | BPF | 2000 | [32] |
| A CW laser + FBGs array | N. A. [b)] | 8 | BPF | 2000 | [33] |
| A broadband optical source + FBGs | ~150 and ~81 | 4 | LPF | 2002 | [34] |
| A broadband optical source + FBGs | ~51 | 8 | LPF | 2005 | [35] |
| Discrete laser array + FBGs | ~588 | 4 | LPF, BPF | 2008 | [36] |
| EO combs | ~10 | 30 | BPF | 2010 | [37] |
| EO combs | ~10 | – [c)] | BPF | 2012 | [38] |
| EO combs | ~25 | 32 | BPF | 2016 | [39] |
| Mode-locked fiber lasers | 5 – 10 | – [c)] | BPF | 2006 | [40] |
| Mode-locked fiber lasers | ~18 | – [c)] | BPF | 2019 | [41] |
| Optical microcombs | ~231 | 21 | BPF | 2014 | [26] |
| Optical microcombs | ~200 | 19 – 21 | LPF, HPF, BPF, BSF | 2018 | [27] |
| Optical microcombs | ~49 | 80 | LPF, BPF | 2019 | [14] |
| Optical microcombs | ~49 | 80 | BPF | 2019 | [15] |
| Optical microcombs | – [d)] | – [c)] | BPF | 2020 | [28] |
| Optical microcombs | ~180 | 8 | BPF | 2022 | [29] |

a) Here we compare the spacings between adjacent wavelength channels in the optical spectra.
b) Here only a single wavelength was employed, and the multiple optical channels were generated by using an optical splitter and an FBGs array.
c) All the generated comb lines were employed without using optical spectral shapers.
d) The channel spacing and tap number were multiplied by triggering a microring resonator with a free spectral range of ~104 GHz to generate a single soliton, two solitons, and soliton crystals.

To achieve high-performance MWP spectral filters using the transversal filtering approach, a large number of wavelength channels are required to serve as discrete taps. In **Table I**, we summarize and compare the MWP transversal filters that have been demonstrated. Although bulky and power-hungry discrete laser arrays [30, 31] and fiber Bragg gratings [12, 32-36] have been used to generate multiple optical channels to supply the needed taps, they suffer from limited available tap numbers due to the significantly increased system size, power consumption, and complexity that come with a large tap number. To provide more taps, laser frequency combs (LFCs) generated by electro-optic (EO) modulation [37-39] and mode-locked fiber lasers have been employed [40, 41], but they still face limitations in terms of operation bandwidth that is intrinsically constrained by their small comb spacings. Recently, optical microcombs [42, 43], which are LFCs generated by micro-scale resonators, offer distinct advantages over traditional multi-wavelength sources for MWP transversal filters, as they can provide a larger number of discrete wavelengths while also having a compact device footprint [20, 22, 44]. Recently, many microcomb-based MWP transversal filters have been demonstrated [14, 15, 27-29], which have realized a variety of filter functions with improved performance and new features compared to conventional MWP transversal filters.

As mentioned, microcomb-based MWP transversal filters can be used for both spectral filtering and temporal signal processing [20, 21, 45, 46]. Recently, we evaluated the performance of temporal signal processors such as differentiators, integrators, and Hilbert transformers [47]. In this paper, we provide a detailed analysis for the performance of MWP spectral filters, specifically low-pass, band-pass, and high-pass filters. Our aim is to explore the impact of several factors on the filter performance, including theoretical limitations resulting from finite tap numbers, experimental component imperfections, and the bandwidths of the signals being filtered. First, we analyze the theoretical limitations on the filter spectral response induced by finite tap numbers, including resolution, roll-off rate, and main-to-secondary sidelobe ratio. Next, we investigate the distortions in the filter response resulting from experimental error sources, including noise of the microcomb, chirp in the EO modulator, third-order dispersion in the dispersive module, shaping errors of the optical spectral shaper, and noise of the photodetector. Finally, we assess the deviations of the filter outputs for input signals with varying bandwidths. These results provide a useful reference for designing and optimizing the performance of MWP spectral filters based on optical microcombs.

## 2. Theory and principle

Microwave transversal filters are widely employed in digital signal processing for a broad range of signal processing applications [45-48]. Implementing them using photonic technologies and hardware can effectively address the electrical bandwidth limitations and provide substantially improved operation bandwidths [21, 47]. The general schematic and processing flow of a MWP transversal filter are shown in **Fig. 1**. To construct such a system, a light source capable of generating multiple wavelengths as discrete taps is required. The input microwave signal is multicast onto all wavelengths by utilizing an EO modulator. Subsequently, a dispersive module introduces time delays between adjacent wavelength channels by delaying the modulated optical replicas. The optical spectral shaping module then assigns specific weights to the delayed replica in each wavelength channel. Finally, the weighted and delayed replicas are combined through photodetection, generating the output microwave signal.

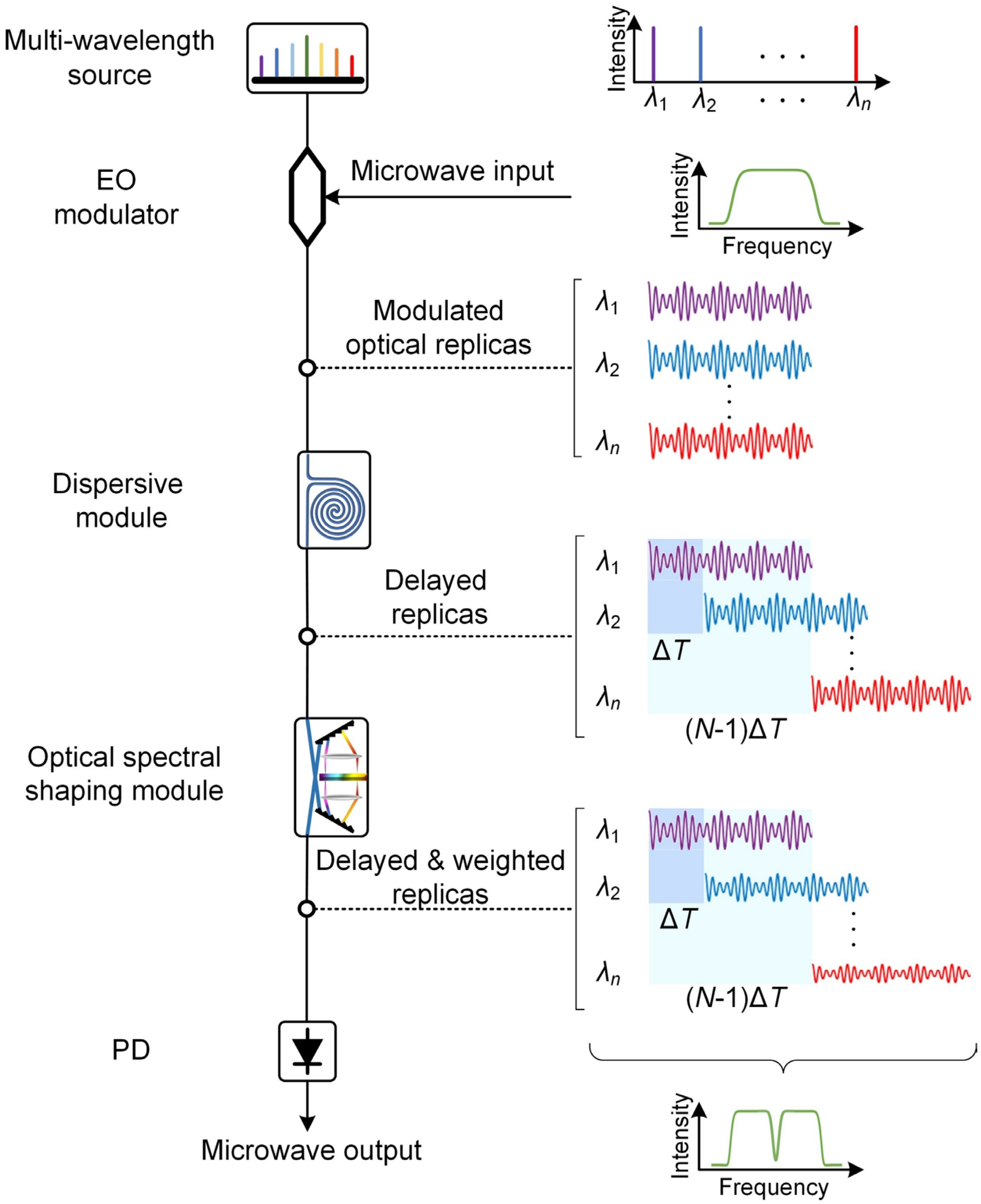

**Fig. 1.** General schematic and processing flow of a microwave photonic (MWP) transversal filter. EO modulator: electro-optic modulator. PD: photodetector.

In order to achieve high performance for MWP transversal filters, a substantial number of discrete wavelengths, or taps, are required. Conventional MWP transversal filters typically implement the multi-wavelength source shown in **Fig. 1** using bulky and power-hungry discrete laser arrays [31, 49], Bragg grating arrays [12, 33], or optical frequency combs generated by EO modulation [50, 51]. These approaches impose significant limitations on the number of taps available for the MWP transversal filter system. In contrast, using a single micro-scale resonator with a high optical nonlinearity and a low loss can simultaneously generate a large number of discrete wavelengths [52], which can bring significant benefits in reducing the system size, complexity, and power consumption [43, 45].

**Fig. 2** illustrates the conceptual diagram for optical microcomb generation in a microring resonator (MRR). The MRR, which has both a high optical nonlinearity and a high quality (Q) factor, is pumped by a high-power continuous-wave (CW) light source near a resonance, resulting in the generation of multiple equally spaced wavelengths at the output. The generation of optical microcombs is based on optical parametric oscillation (OPO), where the parametric gain is provided by the nonlinear optical processes, such as $\chi^{(2)}$ or $\chi^{(3)}$ processes [46, 53, 54]. The cavity where OPO occurs should have both low linear and nonlinear loss to ensure that the gain is sufficient to overcome the cavity loss and sustain the oscillation [55, 56].

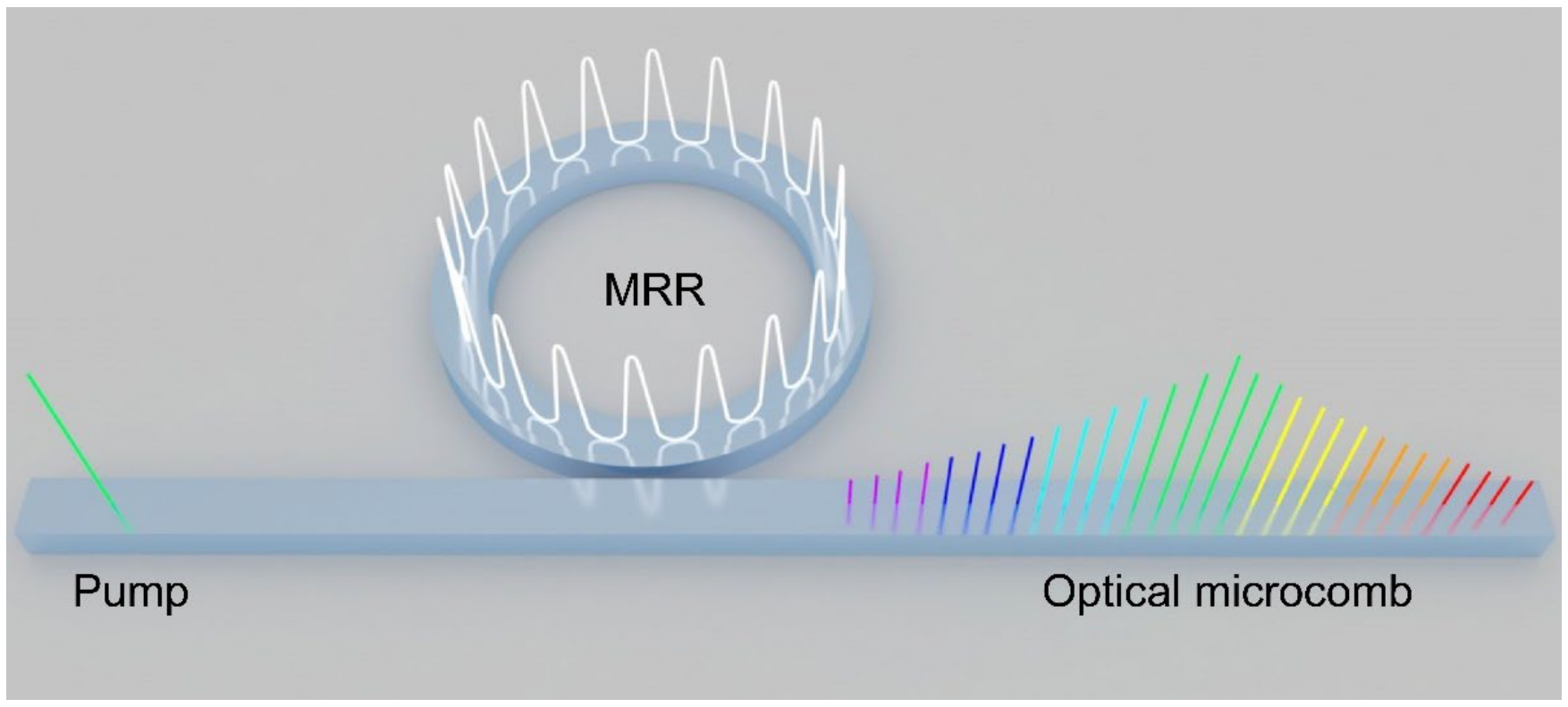

**Fig. 2.** Conceptual diagram of optical microcomb generation in a microring resonator (MRR).

Similar to filters in digital signal processing, microcomb-based MWP transversal filters have a finite impulse response. The spectral transfer function of a MWP transversal filter can be given by [57]

$$H(\omega) = \sum_{n=0}^{M-1} a_n e^{-j\omega n\Delta t} \quad (1)$$

where $M$ is the tap number, $a_n$ ($n$ = 0, 1, 2, …, $M$-1) is the tap coefficient of the $n^{th}$ tap, and $\Delta t$ is the time delay between adjacent wavelength channels, which can be further expressed as

$$\Delta t = \Delta\lambda \times L \times D_2 \tag{2}$$

where $\Delta\lambda$ is the spacing between the adjacent comb lines, $L$ is the fiber length, and $D_2$ is the second-order dispersion (SOD) parameter of the dispersive module. As MWP transversal filters have a finite impulse response, they exhibit a periodic spectral response, and their free spectral range (FSR) can be calculated by:

$$FSR_{MW} = \frac{1}{\Delta t} \tag{3}$$

As can be seen from **Eq. (1)**, by designing the tap coefficients $a_n$ ($n$ = 0, 1, 2, …, $M$-1) through comb shaping, different filter shapes can be achieved without changing the hardware. As a result, microcomb-based MWP transversal filters have a high level of reconfigurability, which makes them well-suited for the implementation of advanced adaptive and dynamic MWP filters.

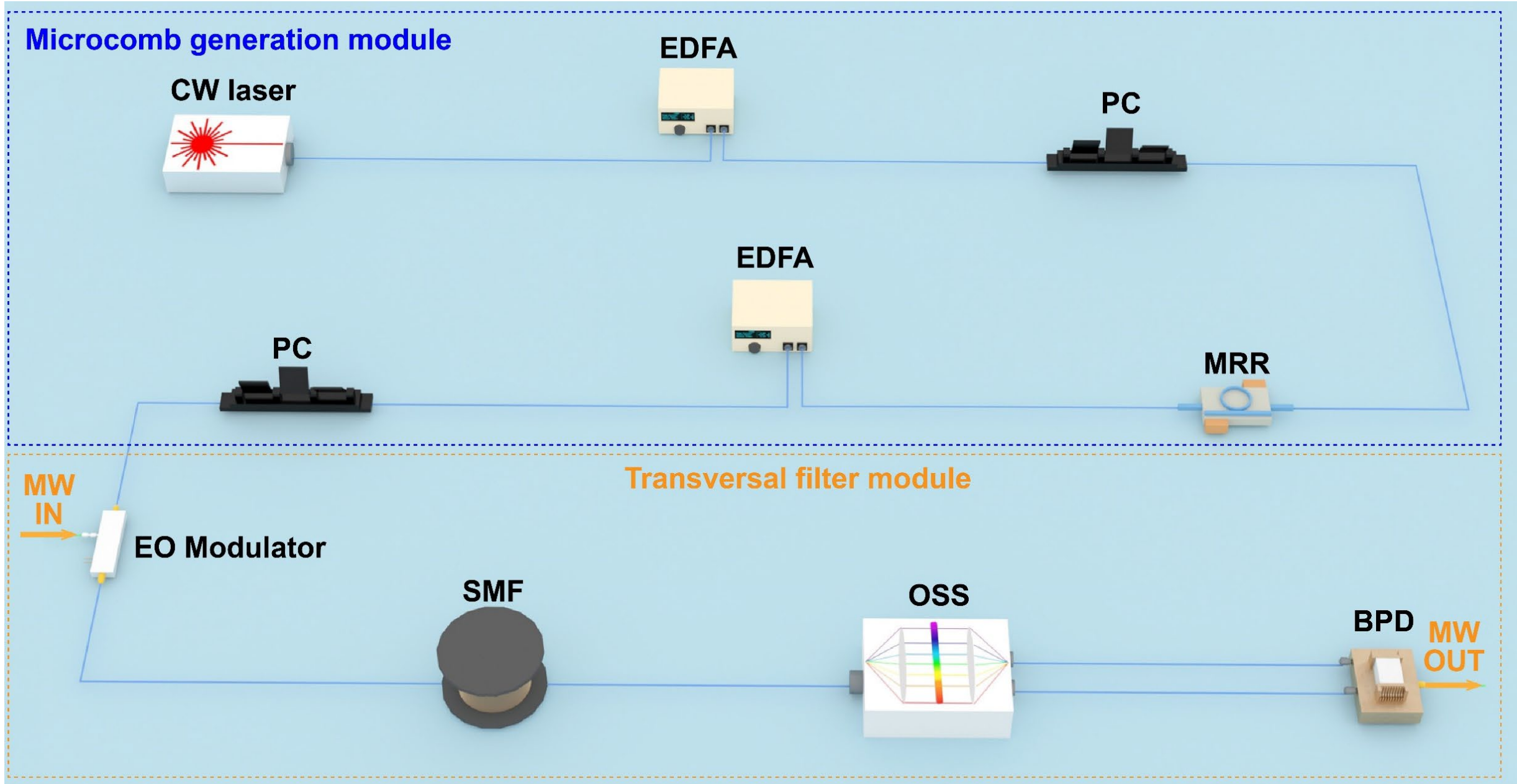

**Fig. 3.** Experimental implementation of a microcomb-based MWP transversal filter. EDFA: erbium-doped fiber amplifier. PC: polarization controller. MRR: micro-ring resonator. OSS: optical spectral shaper. EO modulator: electro-optic modulator. SMF: single-mode fiber. BPD: balanced photo detector. MW: microwave.

**Fig. 3** shows a schematic illustrating the experimental implementation of a microcomb-based MWP transversal filter, which is comprised of two modules, including a microcomb generation module and a transversal filter module. The microcomb generation module generates optical microcombs based on the principle shown in **Fig. 2** to provide multiple equally spaced wavelengths for the subsequent transversal filter module, which performs the processing flow illustrated in **Fig. 1**. A spool of single-mode fiber (SMF) and an optical spectral shaper (OSS) are employed as the dispersive and optical spectral shaping modules in **Fig. 1**, respectively. Two polarization controllers (PCs) are inserted to adjust the polarization of light passing through polarization-sensitive devices like the MRR and EO modulator. To divide all wavelength channels into two groups and introduce a phase difference of π between them, a balanced photodetector (BPD) is used to connect the two complementary output ports of the OSS. This allows for the creation of both positive and negative signs of the tap coefficients.

## 3. Results and discussion

The spectral response of microcomb-based MWP transversal filters in **Eq. (1)** is influenced by both theoretical limitations and experimental system errors. The former is caused by the finite tap number, which limits the filter spectral response with respect to resolution, roll-off rate, and main-to-secondary sidelobe ratio. The latter refers to distortions in the filter response caused by the imperfect response of experimental components in **Fig. 3**, most notably those arising from the noise of microcomb, chirp of the EO modulator, TOD of SMF, OSS shaping errors, and BPD noise. In addition to the limitations in the filters' spectral response mentioned above, in practical signal filtering the system's limited operation bandwidth can also introduce additional errors when filtering wideband signals. In this section, we will provide a thorough analysis of the above factors on the performance of microcomb-based MWP transversal filters. In **Section 3.1**, we will investigate the theoretical limitations in the filter response that arise from finite tap numbers. In **Section 3.2**, we will analyze the distortions in the filter spectral response due to the imperfect response of experimental components. In **Section 3.3**, we will assess the influence of limited operation bandwidth on the experimental performance of the signal filtering. We note that an on-chip microcomb-based MWP transversal filter comprised entirely of integrated components has also been demonstrated recently [29]. Our analysis and discussion also apply to it since it operates on the same principle as the filter in **Fig. 3**.

In our following analysis, we use low-pass filters (LPFs), band-pass filters (BPFs) with $f_{center}$ = 10 GHz, and high-pass filters (HPFs) with a cut-off frequency of 10 GHz as examples to characterize the performance of the microcomb-based MWP transversal filter in **Fig. 3**. These filters are designed based on **Eq. (1)** but with different tap coefficients. The tap coefficients for a LPF can be given by [14]

$$a_{LP,\,n} = 1, \;\; n = 0, 1, 2, \dots, M\text{-}1 \tag{4}$$

which features equal weighting for different taps. Based on the LPF, a BPF can be realized by modifying the tap coefficients as follows [14]

$$a_{BP,\,n} = a_{LP,\,n} \cos\frac{f_{center}\pi n}{FSR_{MW}} e^{-\frac{(n-b)^2}{2\sigma^2}}, \; n = 0, 1, 2, \dots, M\text{-}1 \tag{5}$$

where $f_{center}$ is the center frequency of the filter, and Gaussian apodization is applied to improve the MSSR, with $\sigma$ and $b$ denoting the root mean square width and the peak position of the Gaussian function, respectively.

To achieve an HPF, $R$ sets of tap coefficients corresponding to BPF with increasing center frequencies, given as $a_{BP,n,r}$ ($r$ = 1, 2, …, $R$), are summed up together, which can be expressed as [14]

$$a_{HP,\,n} = \sum_{r=1}^{R} a_{BP,\,n,\,r}, \; n = 0, 1, 2, \dots, M\text{-}1 \tag{6}$$

### *3.1 Theoretical limitations of the filter response*

In this subsection, we analyze the theoretical limitations of microcomb-based MWP transversal filters resulting from finite tap numbers. We focus on evaluating three primary parameters that determine the filter performance, including resolution, roll-off rate (ROR), and main-to-secondary sidelobe ratio (MSSR). The definitions of these three parameters are provided in **Table II** For the analysis in this section, we assume all the components in **Fig. 3** have a perfect response. For comparison, we also assume the filters have the same comb spacing ($\Delta\lambda$ = 0.4 nm, i.e., ~50 GHz) as well as length and dispersion for the SMF ($L$ = 2.1 km and $D_2$ = 17.4 ps/nm/km), which allow for an $FSR_{MW}$ of ~68 GHz.

**Table II Definitions of performance parameters of microcomb-based MWP transversal filters.**

| Parameters | Low-pass filter (LPF) | Band-pass filter (BPF) | High-pass filter (HPF) |
|---|---|---|---|
| Resolution | 3-dB bandwidth of the passband or stopband [a)] | | |
| Roll-off rate (ROR) | The rate at which the filter's amplitude response changes with frequency | | |
| Main-to-secondary sidelobe ratio (MSSR) | The ratio between the amplitude of the main lobe and the first side lobe in the RF amplitude response | | |

a) For the LPF and BPF, the resolution is the 3-dB bandwidth of the passband. For the HPF, the resolution is the 3-dB bandwidth of the stopband.

Although increasing the number of taps can improve the performance for all three types of filters when the comb spacing $\Delta\lambda$ is much larger than the input microwave signal's bandwidth, it is important to note that practical MWP transversal filters are still subject to limitations induced by the number of taps. While microcombs with a large spectral bandwidth can provide a large number of taps, most components in real systems, such as the EDFA and EOM, operate within the telecom C-band (1530 – 1565 nm). Consequently, increasing the number of taps can only be achieved by reducing the comb spacing $\Delta\lambda$. Nevertheless, as per the Nyquist sampling theorem, the comb spacing must be at least twice as wide as the bandwidth of the input RF signal to avoid overlap between modulated RF replicas on different wavelength channels. This implies that a narrow comb spacing would restrict the operation bandwidth of the filter. Thus, for practical applications, one must strike a proper balance between the tap number and comb spacing. In **Fig. 4**, it can be observed that there is only a slight improvement in both the RORs and the MSSRs when the tap number $M$ is increased beyond 80. As a result, $M = 80$ has been widely adopted for microcomb-based MWP transversal filters [14, 15, 27], corresponding to a comb spacing of ~0.4 nm (i.e., ~50 GHz).

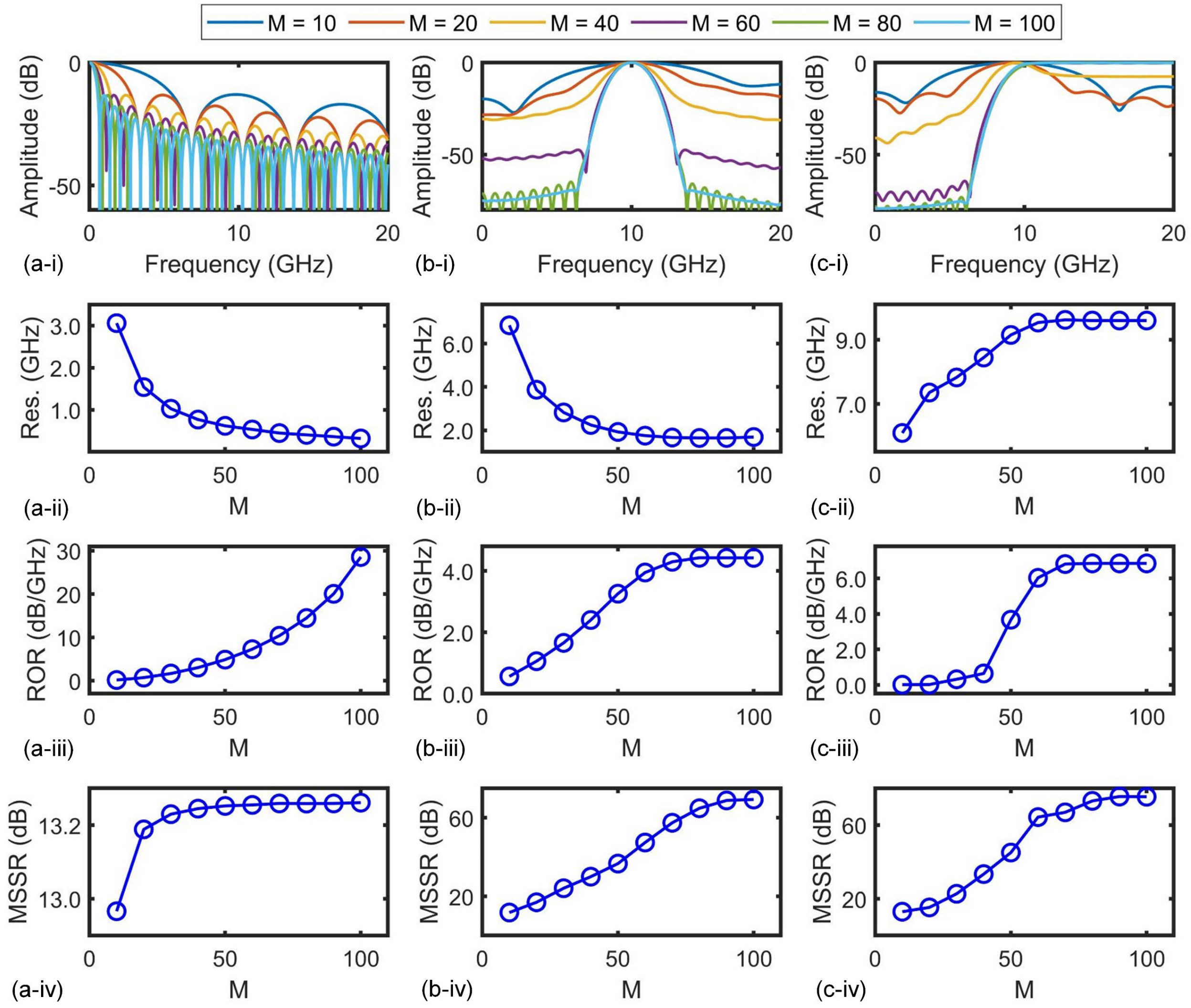

**Fig. 4.** Influence of tap number on performance of (a) low-pass filters (LPFs), (b) band-pass filters (BPFs) with $f_{center}$ = 10 GHz, and (c) high-pass filters (HPFs) passing signals >10 GHz. In (a) – (c), (i) shows RF response of the filters with tap numbers $M$ = 10 – 100, (ii) – (iv) shows resolutions (Res.), roll-off rates (RORs), and main-to-secondary sidelobe ratios (MSSRs) of the filters as a function of $M$, respectively.

### *3.2 Distortions in the filter response induced by imperfect components*

In addition to theoretical limitations, the imperfect response of components in experimentally realized systems gives rise to discrepancies in the filter response. To characterize these, we use average distortion (AD) to describe the difference in the MWP transversal filter's RF amplitude response with and without errors, which is defined as

$$\mathrm{AD} = \sqrt{\sum_{i=1}^{k} \frac{(Y_i - y_i)^2}{k}} \tag{7}$$

where $k$ is the number of sampled points, $Y_1$, $Y_2$, …, $Y_n$ are the RF amplitude response without error sources, and $y_1$, $y_2$, …, $y_n$ are the RF amplitude response with errors induced by imperfect sources. Here, we discuss the influence of different components in real systems on the response of microcomb-based MWP transversal filters. In **Sections 3.2.1** – **3.2.4**, we investigate the influence of specific error sources, assuming the other sources are error-free. In **Section 3.2.5**, we compare the contributions of the different error sources to the overall filter performance. In the analysis of this section, we assume that the tap number $M$ = 80 and all the other parameters are the same as those in **Section 3.1**.

*3.2.1 Influence of microcomb noise*

Microcomb imperfections induce both intensity and phase noise in the wavelength channels. The intensity noise refers to power fluctuations of the comb lines and the intensity noise floor, which mainly arise from photon shot noise and spontaneous emission beat noise [58]. A consequence of imperfect microcombs is that the accuracy of tap coefficients is degraded, leading to a deviation from the ideal filter response that would result from a perfect microcomb.

To characterize the intensity noise of microcombs, the optical signal-to-noise ratio (OSNR) is introduced, which is defined as the ratio of the maximum optical signal power to the noise power in each comb line. **Figs. 5(a-i)** − **(c-i)** show the RF amplitude response of LPFs, BPFs, and HPFs, respectively, with varying OSNR ranging from 10 dB to ∞. The OSNR of ∞ represents the condition of a perfect microcomb with zero intensity noise. As the OSNR increases, the response of all the filters gets closer to that of the filters that would result from a perfect microcomb. We also note that the OSNR has a more significant influence on the response of the BFPs and HPFs.

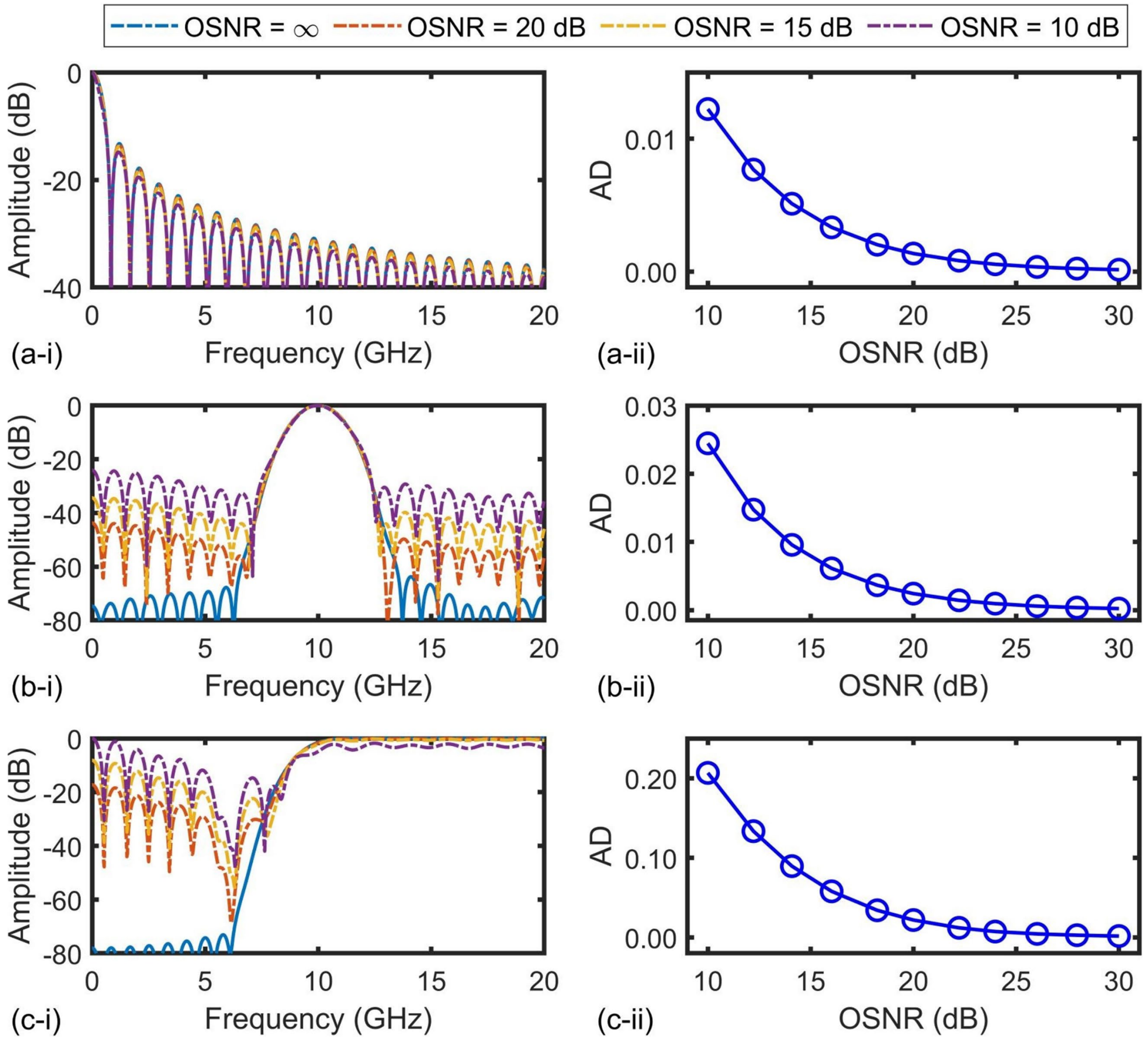

**Fig. 5.** Influence of microcombs' intensity noise on performance of (a) LPFs, (b) BPFs, and (c) HPFs. In (a) – (c), (i) shows RF amplitude response of the filters with different optical signal-to-noise ratios (OSNRs) of the comb lines and (ii) shows average distortions (ADs) from the filter response for a perfect microcomb (with OSNR = ∞) as a function of OSNR.

The ADs for the three types of filters are shown in **Figs. 5(a-ii) − (c-ii)**. As expected, the ADs decrease with OSNR, which agrees with the trend in **Figs. 5(a-i) − (c-i)**. The OSNR induces small distortions (< 0.012) in the response of the LPFs, while giving rise to relatively large distortions for the BPFs and HPFs. The ADs exhibit a sharp decrease when the OSNRs are below 20 dB. As the OSNR increases, the decrease in AD becomes more gradual, and there is only a very small reduction (< 0.005) in AD beyond an OSNR of 20 dB for all the filters.

On the other hand, the phase noise of microcombs, which results in a broadened linewidth, multiple repetition-rate beat notes, and a reduced temporal coherence [59], is difficult to quantitatively analyze. The microcomb's phase noise is influenced by multiple factors, including the noise of the CW pump and the mechanical and thermal noise of the MRR [60, 61]. Using mode-locked microcombs with low phase noise is crucial for achieving stable and long-term operation of microcomb-based MWP transversal filters. This can be accomplished through a variety of mode-locking approaches that have been demonstrated [45, 46].

### *3.2.2 Influence of the EO modulator*

In **Fig. 3**, an EO modulator is employed to modulate the input microwave signal onto each wavelength channel. Due to their high modulation efficiency, large operation bandwidth, and low insertion loss, Mach-Zehnder modulators (MZMs) are commonly used [62]. In addition to intensity modulation, as the existence of an asymmetry in the electric field overlap at each electrode [63], experimental MZMs also result in undesired phase modulation, namely modulation chirp. In this subsection, we analyze the influence of modulation chirp on the filter performance. The chirp can be characterized by the chirp parameter defined as [64]

$$\alpha = \frac{\gamma_1+\gamma_2}{\gamma_1-\gamma_2} \tag{8}$$

where $\gamma_1$ and $\gamma_2$ are the voltage-to-phase conversion coefficients for the two arms of the MZM. When $\alpha = 0$ (i.e., $\gamma_1 = -\gamma_2$), pure intensity modulation is achieved, whereas for $\alpha = \infty$ (i.e., $\gamma_1 = \gamma_2$), it is pure phase modulation. Modulation chirp causes distortions in the optical signals after modulation, thus leading to distortions in the response of the filters.

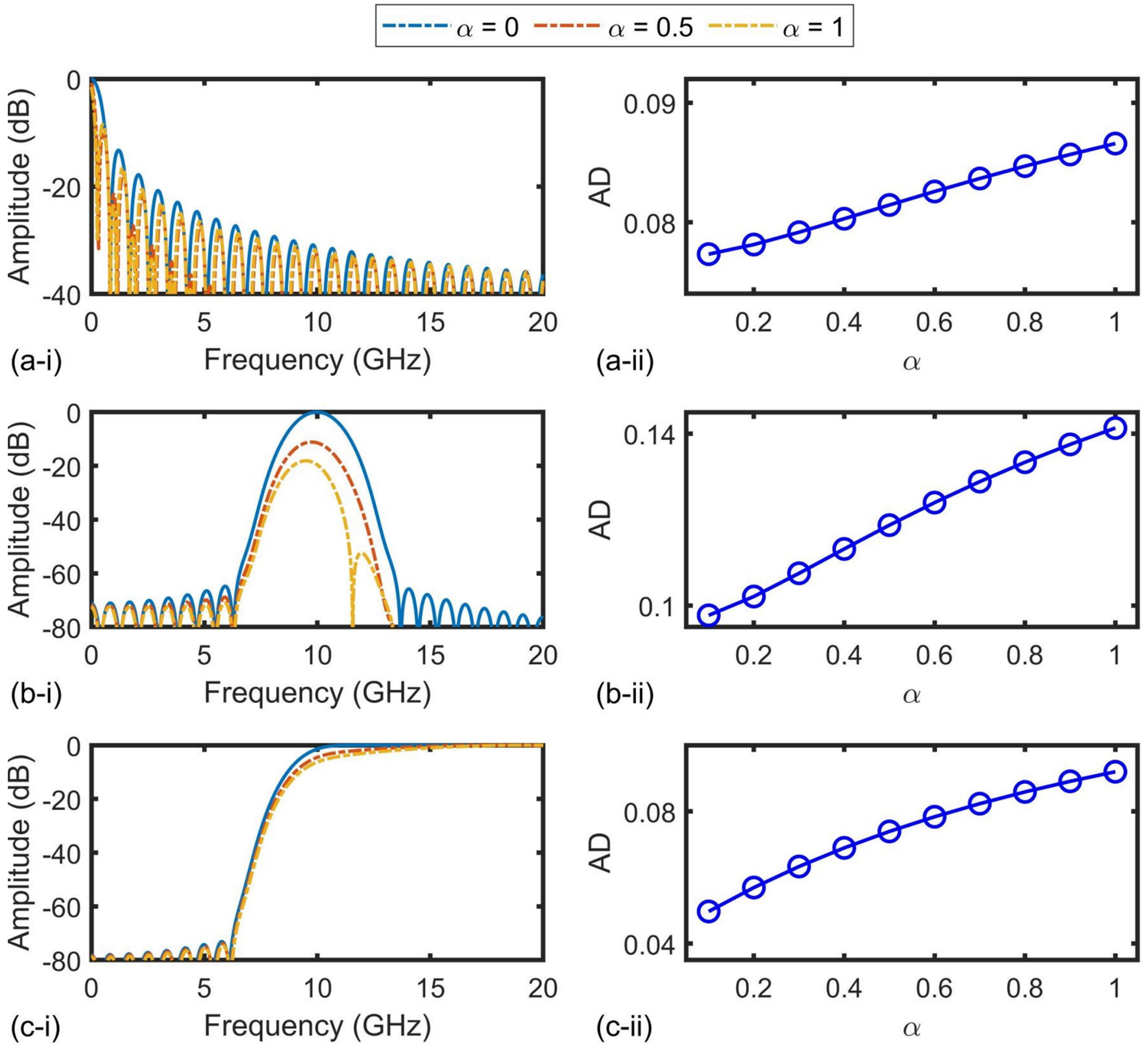

**Fig. 6.** Influence of modulation chirp on performance of (a) LPFs, (b) BPFs, and (c) HPFs. In (a) – (c), (i) shows RF amplitude response of the filters for different chirp parameter $\alpha$ and (ii) shows ADs from filter response for chirp-free modulation (with $\alpha = 0$) as a function of $\alpha$.

**Figs. 6(a-i) − (c-i)** show the RF amplitude response of LPFs, BPFs, and HPFs, respectively, with varying chirp parameters $\alpha$ ranging from 0 to 1. The $\alpha = 0$ corresponds to the condition of chirp-free modulation. As $\alpha$ decreases, the response of all the filters approaches that of the filter with chirp-free modulation, indicating less distortion caused by a low modulator chirp. The calculated ADs as a function of $\alpha$ are shown in **Figs. 6(a-ii) − (c-ii)**, where the ADs increase with $\alpha$, showing agreement with the trend in **Figs. 6(a-i) − (c-i)**. The impact of modulation chirp on the filter response is more pronounced for the BPFs than for the LPFs and HPFs. It is worth noting

that the modulation chirp can result in various types of distortions in the filter response, depending on the type of filter. For the LPFs, although the ROR is increased from 10.76 dB/GHz to 38.12 dB/GHz, the MSSR is degraded from 13.26 to 6.96. For the BPFs, the ROR is almost unchanged, but the MSSR is decreased from 64.79 to 52.9. For the HPFs, the ROR is decreased from 7.0 dB/GHz to 0.67 dB/GHz, and the MSSR is almost unchanged.

#### *3.2.3 Influence of the SMF*

In **Fig. 3**, A spool of SMF is used as the dispersive module to introduce time delays between adjacent wavelength channels. The second-order dispersion (SOD) of the SMF is desired to produce uniform time delays, while the existence of third-order dispersion (TOD) introduces non-uniform time delays, hence giving rise to undesired phase errors. In this subsection, we analyze the distortions induced by TOD of the SMF.

The additional non-uniform time delays of the $n^{\text{th}}$ tap induced by TOD can be expressed as [44]

$$\Delta T_{TOD} = D_3 \, L \, \Delta\lambda^2 \, n^2 \tag{9}$$

where $D_3$ is the TOD parameter. **Figs. 7(a-i) − (c-i)** show the RF amplitude response of LPFs, BPFs, and HPFs, respectively, with varying $D_3$ ranging from 0 to 0.2 ps/nm$^2$/km. The $D_3 = 0$ corresponds to the condition of an SMF with zero TOD. For all the filters, as $D_3$ decreases, their response gets closer to that corresponding to the SMF with zero TOD. This indicates decreased distortions in the filter response for a smaller TOD parameter. When $D_3$ increases from 0 to 0.2 ps/nm$^2$/km, the MSSR for the LPFs is slightly degraded from 13.26 to 13.06. For the BPFs and HPFs, the RORs are decreased from 4.42 dB/GHz to 3.85 dB/GHz and from 7.0 dB/GHz to 6.39 dB/GHz, respectively.

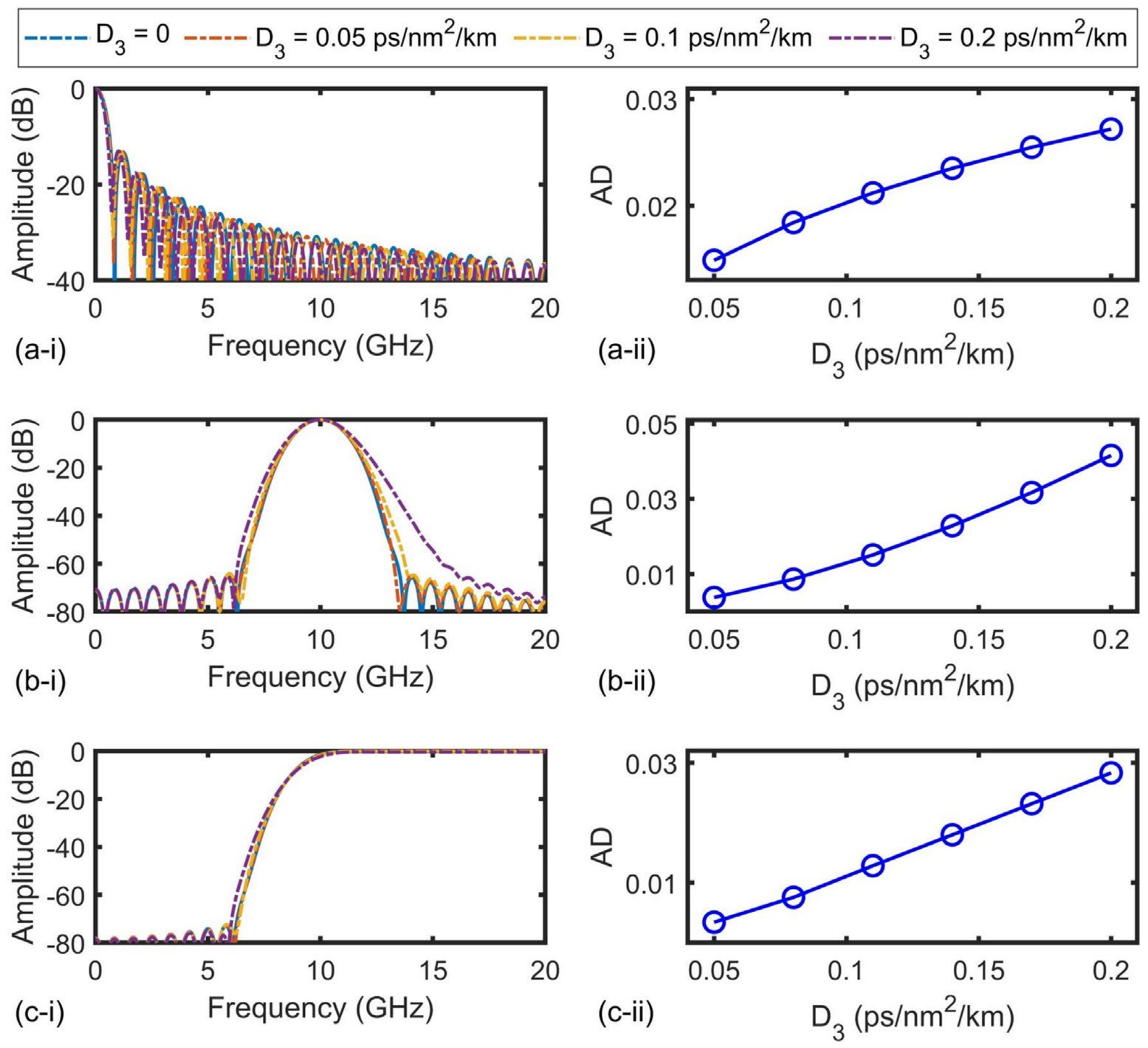

**Fig. 7.** Influence of SMF's third-order dispersion (TOD) on performance of (a) LPFs, (b) BPFs, and (c) HPFs. In (a) – (c), (i) shows RF response of the filters for different TOD parameters $D_3$ and (ii) shows ADs from the filter response for an SMF with zero TOD (with $D_3 = 0$) as a function of $D_3$.

**Figs. 7(a-ii) − (c-ii)** show the ADs versus $D_3$, where the ADs increase with $D_3$ for all the filters, showing agreement with the trend in **Figs. 7(a-i) − (c-i)**. We also note that the influence of the TOD on the BPFs' response is more significant as compared to the LPFs and HPFs, indicating that the BPFs require a higher level of phase accuracy among different wavelength channels.

### *3.2.4 Influence of the optical spectral shapers and BPDs*

Here we analyze the distortions induced by the OSS and BPD in **Fig. 3**. The OSS is employed to apply the designed tap weights to the delayed signals across different wavelength channels, and the BPD is used to sum the delayed and weighted signals and generate the microwave output. The OSS can induce shaping errors, which give rise to inaccurate tap weights, consequently resulting in distortions in the filter response. In addition, the distortions can be caused by the presence of noise and uneven transmission response of the BPD.

We introduce the concept of shaping error range, which refers to the random tap weight shaping errors within a certain percentage range, to characterize the shaping errors of the OSS. The RF response of all filters is shown in **Figs. 8(a-i)** − **(c-i)**, respectively, with different shaping error ranges varying from 0% to 10%. A shaping error range of 0% represents the condition of an OSS without any shaping errors. As the shaping error range increases, the response for all the filters deviates further from an error-free OSS, suggesting an increase in the distortions caused by larger shaping errors of the OSS. When the shaping error range increases from 0% to 10%, for the LPFs, the ROR and MSSR have shown negligible differences. For the BPFs, the MSSR decreases from 64.79 to 32.15. For the HPFs, the MSSR reduces from 73.1 to 64.79. When shaping error range increases from 0 to 10%, the ROR and MSSR for the LPFs show insignificant differences. However, for the BPFs, the MSSR decreases from 64.79 to 32.15, while for the HPFs, the MSSR decreases from 73.1 to 64.79.

**Figs. 8(a-ii)** − **(c-ii)** show the ADs as a function of the shaping error range. The ADs increase with the shaping error range for all the filters, which agrees with the trend in **Figs. 8(a-i)** − **(c-i)**. Moreover, it is worth noting that the impact of OSS shaping errors on the response of the HPFs is more significant compared to that of the LPFs and BPFs, reflecting that the HPFs require a higher level of shaping accuracy.

Through using a BPD, the common-mode noise of optical signals is effectively suppressed. Therefore, the distortions induced by the BPD mainly stem from its limited response bandwidth and uneven transmission response, which causes additional tap weight errors after spectral shaping. Similar to the BPD, the EO modulator's limited response bandwidth and uneven transmission response also give rise to tap weight errors. These tap weight errors and the OSSs' shaping errors can be effectively alleviated via feedback control. We also note that, shot noise of the BPD introduces stochastic power fluctuations, which constrains the lowest achievable phase noise floor [65]. Distortions induced by this are similar to those induced by the phase noise of the microcomb and can be reduced by using a BPD with an improved sensitivity [66].

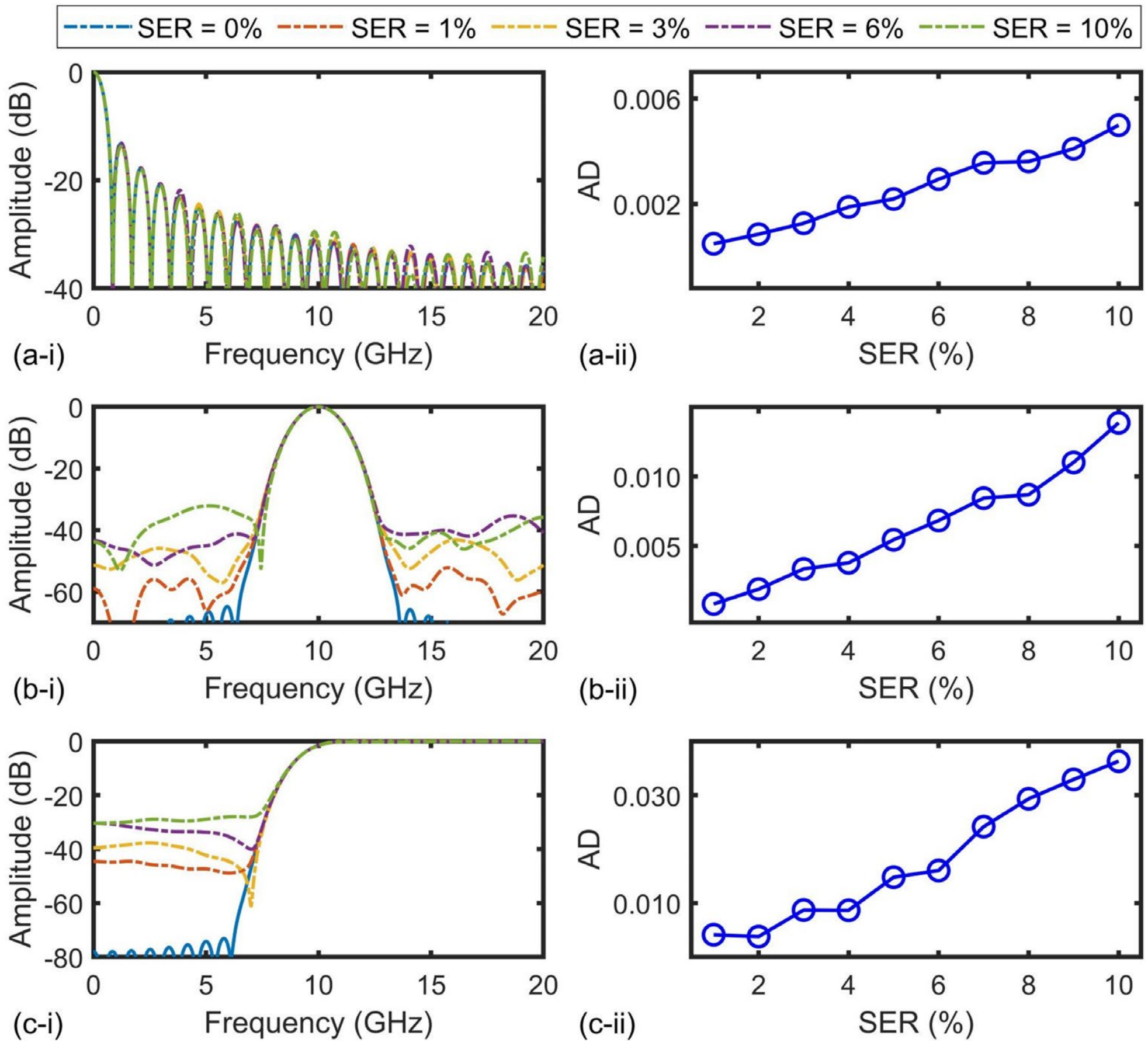

**Fig. 8.** Influence of OSS shaping errors on performance of (a) LPFs, (b) BPFs, and (c) HPFs. In (a) – (c), (i) shows RF response of the filters for different shaping error ranges and (ii) shows ADs from the filter response for an OSS without any shaping errors (i.e., with a shaping error range = 0%) as a function of shaping error range.

### *3.2.5 Contributions of different error sources*

Here, we analyze the contributions of the error sources discussed above to the overall distortions in the response of microcomb-based MWP transversal filters. **Figs. 9(a)** – **(c)** show the simulated RF amplitude response for all three types of filters, after accounting for the distortions induced by different error sources including the (I) OSNR of microcombs, (II) chirp of EO modulator, (III) TOD of SMF, and (IV) errors of the OSS and BPD. The ideal filter response without any errors is also shown for comparison. Based on the measured parameters of the components in our previous experiments [17, 22, 67], the chirp parameter of the EO modulator, the TOD parameter of the SMF, and the random shaping error range are set to $\alpha = 0.5$, $D_3 = 0.083$ ps/nm$^2$/km, and shaping error range = 5%, respectively. In our simulations, we also used the OSNRs of the comb lines of a practical microcomb that were measured by an optical spectrum analyzer. Consistent with expectations, the overall distortions in the filter response exhibit a noticeable increase as a consequence of the accumulation of the errors induced by the imperfect response of components mentioned above.

To quantify the contributions of different error sources, we calculate the ADs from the results in **Figs. 9(a)** − **(c)** and plot them in **Fig. 9(d)**. The ADs increase with the accumulation of the errors induced by sources I – IV, showing agreement with the trend in **Figs. 9(a)** − **(c)**. For all the three filters, the main source of distortions is the EO modulator chirp. Compared to the LPF and HPF, the distortions of the BPF are more significantly influenced by the TOD of the SMF. In addition, the distortions caused by the imperfect microcomb affect the HPF more significantly than the LPF and BPF.

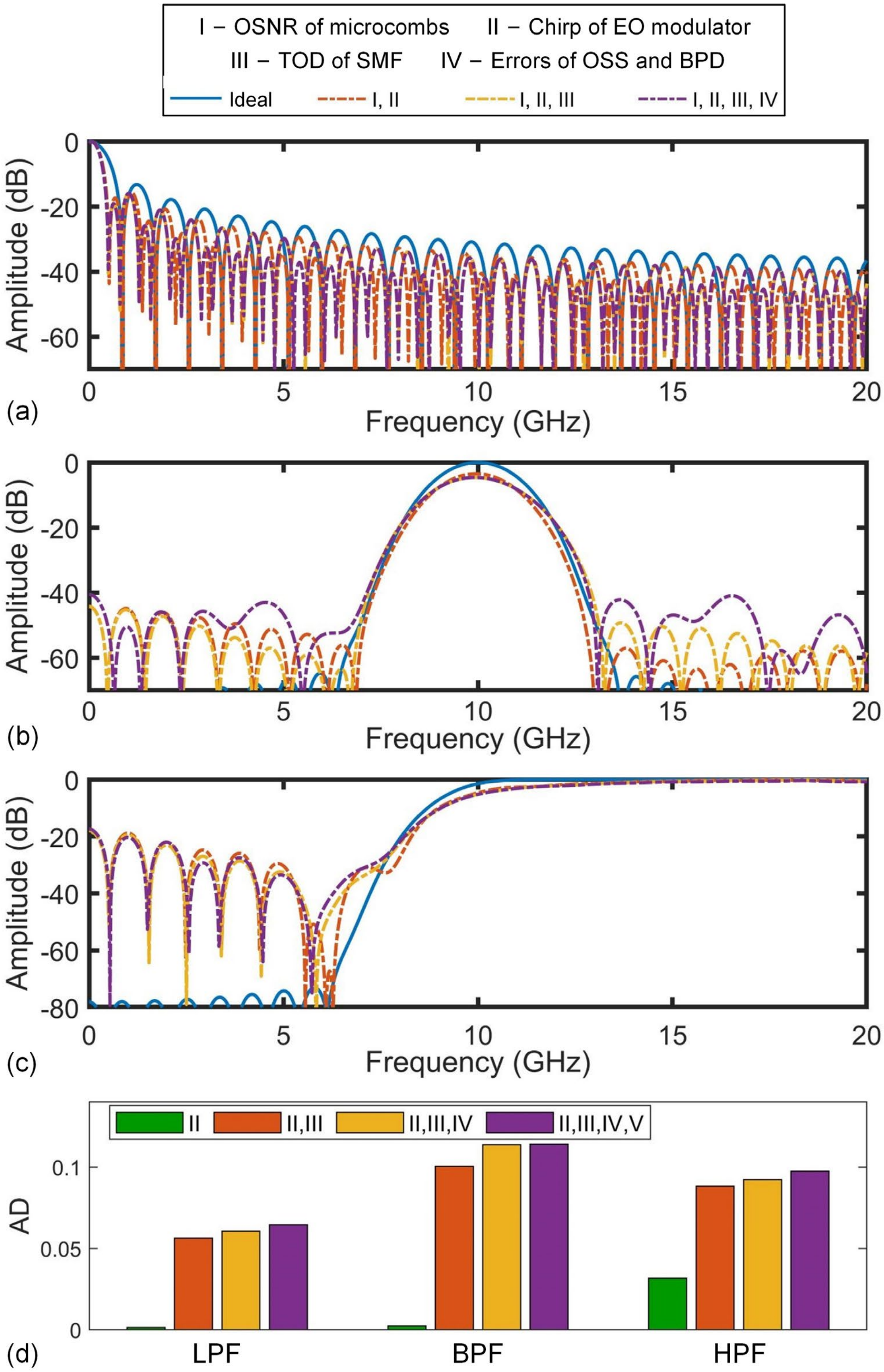

**Fig. 9.** Contributions of different error sources to the overall distortions of (a) LPFs, (b) BPFs, and (c) HPFs. Different curves show the RF amplitude response after accumulating errors induced by different sources from I to IV. The ideal filter response without any errors is also shown for comparison. (d) Corresponding ADs for the three filters after accumulating errors induced by different sources from I to IV.

### *3.3 Influence of signal bandwidth*

For signal filtering, according to the theory of signals and systems [68], the ideal filter output can be given by

$$S_{out}(\omega) = S_{in}(\omega) \cdot H(\omega) \tag{10}$$

where $S_{in}(\omega)$ and $S_{out}(\omega)$ are the input and output microwave signals in the spectral domain, respectively, and $H(\omega)$ is the spectral transfer function in **Eq. (1)**. In **Sections 3.1** and **3.2**, we have analyzed the influences of both theoretical limitations and experimental system errors on $H(\omega)$. In practical signal filtering, the system's limited operation bandwidth introduces additional errors when filtering wideband signals, thus leading to deviations between the filter output signal and the ideal output signal in **Eq. (10)**. In this section, we analyze the errors introduced in microcomb-based MWP transversal filters when processing microwave signals with varying spectral bandwidths.

As discussed in **Section 3.2**, a microcomb-based MWP transversal filter has a periodic spectral response due to its finite impulse response. The FSR of the spectral response is equal to the inverse of the time delay between adjacent wavelength channels. On the other hand, according to the Nyquist sampling theorem, a continuous-time signal that is bandwidth-limited requires sampling at a rate greater than twice its maximum frequency component to prevent aliasing. This limitation sets a maximum allowable bandwidth for the RF signal to be processed, which should not exceed half of the microcomb's comb spacing. Therefore, the operation bandwidth of a microcomb-based MWP transversal filter can be expressed as:

$$OBW = \min \{\Delta\lambda/2, FSR_{MW}/2\} \tag{11}$$

where $\Delta\lambda$ is the comb spacing, $FSR_{MW}$ is the FSR of the RF response in **Eq. (3)**, and min {-} represents taking the minimum value between the two.

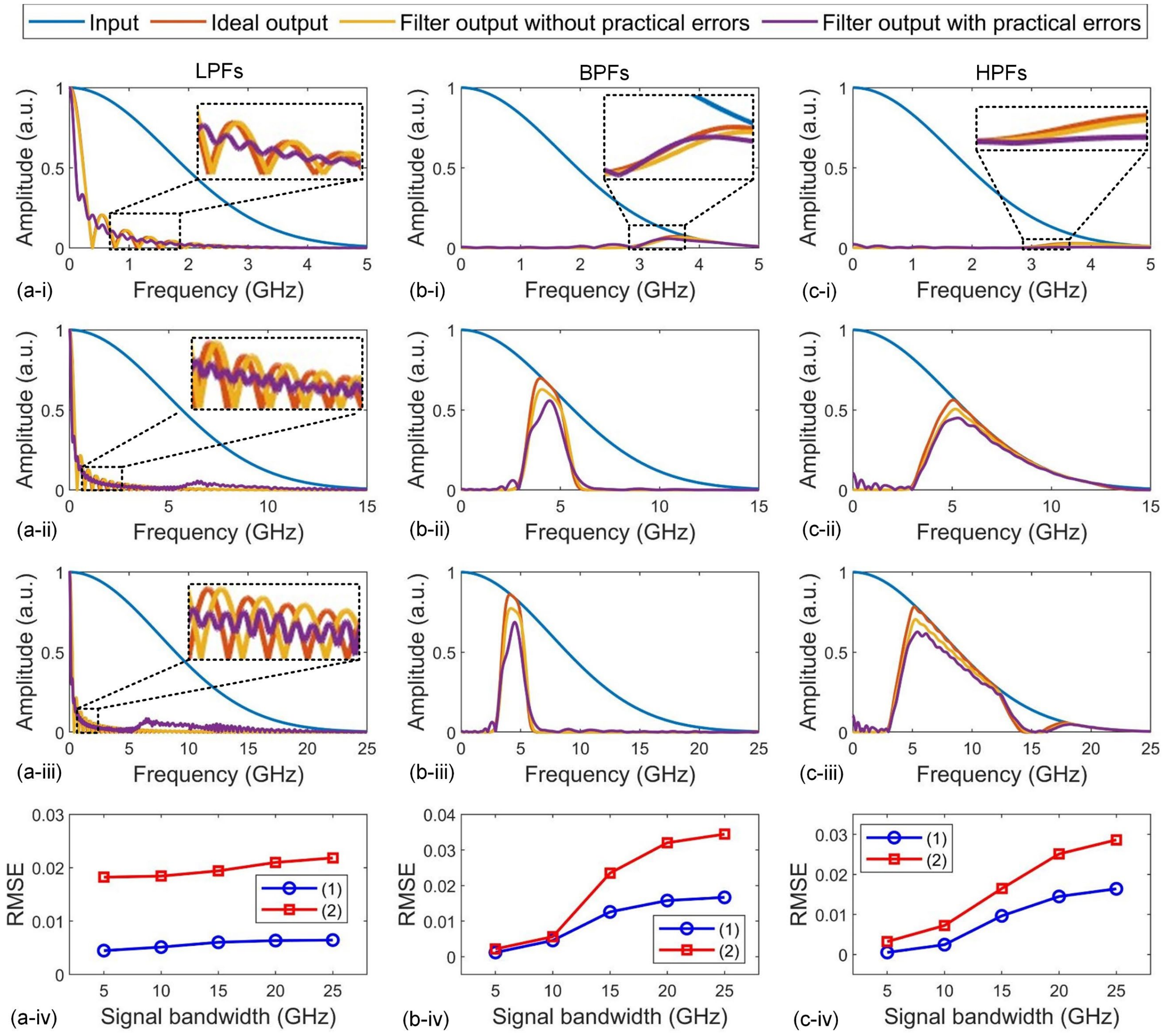

**Fig. 10.** Filter outputs of input Gaussian pulses with different spectral bandwidth for (a) LPFs, (b) BPFs, and (c) HPFs. In (a) – (c), (i) – (iii) show the spectra of input Gaussian pulses with spectral bandwidth of (i) ~5 GHz, (ii) ~15 GHz, and (iii) ~25 GHz, and corresponding filter outputs with and without experimental errors. The ideal outputs are also shown for comparison. (iv) shows the corresponding RMSEs of the filter outputs as a function of input signal bandwidth, where curves (1) and (2) show the results for the filter outputs without and with experimental errors, respectively.

Based on the physical processes of signal delay and summation in the transversal filter, assuming all the components in **Fig. 3** are error-free, the filter output can be expressed as

$$S_{out}(\omega) = \mathrm{IFT}[s_{out}(t)] = \mathrm{IFT}[\sum_{n=0}^{M\text{-}1} a_n s_{in}(t - n\Delta t)] \tag{12}$$

where $s_{out}(t)$ and $S_{out}(\omega)$ are the output microwave signal in the time and spectral domain, respectively, $s_{in}(t)$ is the input microwave signal in the time domain, and IFT[-] denotes the operation of the inverse Fourier transform. In **Eq. (12)**, $a_{1,2,...,M-1}$ and $\Delta t$ are the designed tap coefficients and time delay in **Eq. (1)**, respectively. Due to the limited operation bandwidth of an experimental system, the output signal calculated from **Eq. (12)** differs from that calculated from **Eq. (10)**. After accounting for the influence of the imperfect response of real components on the tap coefficients and time delay, the filter output with experimental errors can be given by

$$S_{out}'(\omega) = \mathrm{IFT}[s_{out}'(t)] = \mathrm{IFT}[\sum_{n=0}^{M-1} a'_n s_{in}(t - n\Delta t')] \tag{13}$$

where $s_{out}'(t)$ and $S_{out}'(\omega)$ are the experimental output microwave signals in the time and spectral domain, respectively, $a'_{1,2,...,M-1}$ and $\Delta t'$ are the tap coefficients and time delay with errors induced by the imperfect response, respectively.

To quantify the errors in the filter outputs, the root mean square errors (RMSE) is introduced. Similar to the definition of AD, the RMSE is defined as [45]

$$\mathrm{RMSE} = \sqrt{\sum_{r=1}^{R} \frac{(Y_r - y_r)^2}{R}} \tag{14}$$

where $R$ is the number of sampled points, $Y_1$, $Y_2$, …, $Y_r$ are the values of ideal output signal spectra calculated based on **Eq. (10)**, and $y_1$, $y_2$, …, $y_r$ are the values of the filter output spectra calculated based on either **Eq. (12)** or **Eq. (13)**. **Figs. 10(a-iv) − (c-iv)** show the calculated RMSEs for the filter outputs as a function of the input signal bandwidth. As expected, the RMSEs increase with the spectral bandwidth of the input signal for all filters, and the RMSEs for the filter outputs with experimental errors are higher than those for the filter outputs without experimental errors. It should also be noted that when the spectral bandwidth of the input signal is larger than $FSR_{MW}$ / 2 = 15 GHz but lower than $\Delta\lambda$ / 2 = 25 GHz, the filter can still work but with notable inaccuracies.

However, if the spectral bandwidth of the input signal increases beyond 25 GHz, the filter cannot function properly. This is because there will be aliasing between the modulated signals of adjacent wavelength channels due to the comb spacing being 50 GHz. This work has wide applicability to microwave photonic devices [69 - 96] based on optical microcombs [97 - 119] with potential applications to quantum optical devices [120 - 132] as well.

## 4. Conclusions

In summary, we present a detailed analysis for the performance of microcomb-based MWP spectral filters achieved by employing a transversal filter approach, and determine how they are affected by theoretical limitations due to finite tap numbers, imperfect response of experimental components, and bandwidths of input microwave signals. We first investigate the theoretical limitations caused by finite tap numbers on the filter response, focusing on the filter parameters such as resolution, ROR, and MSSR. Next, we analyze the distortions in the filter response resulting from the various experimental error sources, including noise of the microcomb, chirp in the EO modulator, TOD in the dispersive module, shaping errors of the optical spectral shaper, and noise of the photodetector. Finally, we investigate the influence of input signal bandwidth on the filtering errors. We investigate three types of filters including the LPFs, BPFs, and HPFs. Our results show that the influence of different factors varies for distinct filter shapes studied here. This work provides a useful guide for optimizing the performance of microcomb-based MWP transversal filters, which offer a high reconfigurability and enable the realization of diverse filter functions without changing the hardware.

## Declaration of competing interest

The authors declare no competing interests.

## Data availability

The authors declare that the data supporting the findings of this study are available.